\input amstex
\magnification 1100
\TagsOnRight
\def\qed{\ifhmode\unskip\nobreak\fi\ifmmode\ifinner\else
 \hskip5pt\fi\fi\hbox{\hskip5pt\vrule width4pt
 height6pt depth1.5pt\hskip1pt}}
\NoBlackBoxes
\baselineskip 19 pt
\parskip 5 pt
\def\stretch {\noalign{\medskip}}
\define \bC {\bold C}
\define \bCp {\bold C^+}
\define \bCm {\bold C^-}
\define \bCpb {\overline{\bold C^+}}
\define \ds {\displaystyle}
\define \bR {\bold R}
\define \bm {\bmatrix}
\define \endbm {\endbmatrix}

\centerline {\bf FACTORIZATION OF THE TRANSITION MATRIX}
\centerline {\bf FOR THE GENERAL JACOBI SYSTEM}

\vskip 20 pt
\centerline {Tuncay Aktosun}
\vskip -3 pt
\centerline {Department of Mathematics}
\vskip -3 pt
\centerline {University of Texas at Arlington}
\vskip -3 pt
\centerline {Arlington, TX 76019-0408, USA}
\vskip -3 pt
\centerline {aktosun\@uta.edu}

\vskip 10 pt
\centerline {Abdon E. Choque-Rivero}
\vskip -3 pt
\centerline {Instituto de F\'{\i}sica y Matem\'aticas}
\vskip -3 pt
\centerline {Universidad Michoacana de San Nicol\'as de Hidalgo}
\vskip -3 pt
\centerline {Ciudad Universitaria, C.P. 58048}
\vskip -3 pt
\centerline {Morelia, Michoac\'an, Mexico}
\vskip -3 pt
\centerline {abdon\@ifm.umich.mx}

\vskip 15 pt

\noindent {\bf Abstract}:
The Jacobi system on a full-line lattice is considered when it contains additional
weight factors.
A factorization formula is derived expressing the scattering from
such a generalized Jacobi system in terms of the scattering from its fragments.
This is done by writing the transition matrix for the generalized
Jacobi system as an ordered matrix product of the transition matrices
corresponding to its fragments. The resulting factorization formula
resembles the factorization formula for the Schr\"odinger equation
on the full line.

\vskip -5 pt
\par\noindent {\bf Short title:} Factorization for the Jacobi system
\vskip -5 pt
\par\noindent {\bf Keywords:} Jacobi system,
discrete systems, scattering,
scattering from fragments, transition matrix, factorization formula

\newpage

\noindent {\bf 1. INTRODUCTION}
\vskip 3 pt

Consider the Schr\"odinger equation on the full line $\bR:=(-\infty,+\infty)$
given by
$$-\psi''(k,x)+V(x)\, \psi(k,x)=k^2\,\psi(k,x),\qquad x\in\bR,
\tag 1.1$$
where the prime denotes the $x$-derivative and the potential
$V$ is real valued and satisfies
$$\ds\int_{-\infty}^\infty dx\, (1+|x|)\,|V(x)| <+\infty.\tag 1.2$$
The restriction given in (1.2) allows us [10-13,20,24] to
develop a mathematical theory of scattering for (1.1), assures the
existence of scattering solutions, and also guarantees that there are
at most a finite number of bound states.

There are two particular solutions to (1.1). The first is the Jost solution from
the left, denoted by $f_{\text l}(k,x),$ satisfying the spatial asymptotics
$$f_{\text l}(k,x)=e^{ikx}\left[1+o(1)\right],\qquad x\to+\infty,\tag 1.3$$
and the other is the Jost solution from
the right, denoted by $f_{\text r}(k,x),$ satisfying the spatial asymptotics
$$f_{\text r}(k,x)=e^{-ikx}\left[1+o(1)\right],\qquad x\to-\infty.\tag 1.4$$
The scattering coefficients corresponding to $V$ are obtained
through the spatial asymptotics
$$f_{\text l}(k,x)=\ds\frac{1}{T(k)}\,e^{ikx}+\ds\frac{L(k)}{T(k)}\,e^{-ikx}+o(1),\qquad x\to-\infty,\tag 1.5$$
$$f_{\text r}(k,x)=\ds\frac{1}{T(k)}\,e^{-ikx}+\ds\frac{R(k)}{T(k)}\,e^{ikx}+o(1),\qquad x\to+\infty,\tag 1.6$$
where $T$ is the transmission coefficient,
$R$ is the reflection coefficient from the right, and $L$ is the reflection coefficient from the left. Corresponding to the potential $V,$
let us use $\Lambda(k)$ to denote the $2\times 2$
transition matrix defined for $k\in\bR$ as
$$\Lambda(k):=\bm \ds\frac{1}{T(k)}& -\ds\frac{R(k)}{T(k)}\\
\stretch
\ds\frac{L(k)}{T(k)}& \ds\frac{1}{T(-k)}\endbm.\tag 1.7$$

Let us fragment the real axis $\bold R$ into $N$ pieces as
$$\bR=(x_0,x_1]\cup (x_0,x_1]\cup (x_1,x_2] \cup \cdots  \cup (x_{N-1},x_N],$$
where $N\ge 2$ and we have
$$x_0:=-\infty,\quad x_1<x_2<\cdots < x_{N-1},\quad x_N:=+\infty.$$
We then obtain the fragmentation of the potential $V$ into $N$ pieces as
$$V(x)=\cases V_1(x),\qquad & x\in (x_0,x_1],\\
V_2(x),\qquad & x\in (x_1,x_2],\\
\quad \vdots & \quad \vdots \\
V_N(x),\qquad & x\in (x_{N-1},x_N],\endcases\tag 1.8$$
where for $j=1,2,\dots,N$ we have defined
$$V_j(x):=\cases V(x),&\qquad x\in (x_{j-1},x_j],\\
0,& \qquad x\not\in (x_{j-1},x_j].\endcases$$

Let us use $f_{{\text l}j}(k,x)$ and $f_{{\text r}j}(k,x)$ to denote the
Jost solutions from the left and from the right, respectively,
corresponding to the potential $V_j.$ Let us also use
$T_j,$ $R_j,$ $L_j$ for the respective scattering coefficients corresponding
to $V_j.$ Thus, $f_{{\text l}j}(k,x)$ and $f_{{\text r}j}(k,x)$ satisfy
(1.1) when $V$ there is replaced with $V_j,$
and they satisfy (1.3) and (1.4), respectively.
They also satisfy (1.5) and (1.6), respectively, with the scattering
coefficients $T,$ $R,$ $L$ replaced with $T_j,$ $R_j,$ $L_j,$
respectively.
Let $\Lambda_j(k)$ be the corresponding transition matrix
for $V_j$ defined in a similar way as in (1.7) as
$$\Lambda_j(k):=\bm \ds\frac{1}{T_j(k)}& -\ds\frac{R_j(k)}{T_j(k)}\\
\stretch
\ds\frac{L_j(k)}{T_j(k)}& \ds\frac{1}{T_j(-k)}\endbm.$$
Corresponding to the fragmentation (1.8), we have the factorization formula
[2,4,17,19,25-27]
$$\Lambda(k)=\Lambda_1(k)\, \Lambda_2(k)\, \cdots \Lambda_N(k),
\qquad k\in\bR,\tag 1.9$$
where the right-hand side consists of the product of the $2\times 2$
transition matrices
in the order indicated.

The factorization formula (1.9) and its various generalizations [2-4] allow us to
understand how the scattering from a system develops in terms of the scattering
from the components of that system. It has important applications in various
areas, such as determining the phase [5,7] of a complex-valued reflection coefficient
based on amplitude measurements,
determining material properties of thin films via neutron reflectometry [6,21-23],
quantum wires [17,18], quantum computing [18], and quantum
scattering from coupled systems [16].

Our goal in this paper is to derive the analog of the factorization formula
(1.9)
for some difference equations related to
(1.1) and its generalizations. In particular, we will
do so for the general Jacobi system
on the full-line lattice given by [9]
$$a(n+1)\,\psi(\lambda,n+1)+b(n)\,\psi(\lambda,n)+
a(n)\,\psi(\lambda,n-1)=\lambda\,w(n)\,\psi(\lambda,n),
\qquad n\in\bold Z,\tag 1.10$$
where $\bold Z$ denotes the set of integers,
$\lambda$ is the spectral parameter, and
$a(n),$ $b(n),$ $w(n)$ are some real coefficients that
may depend on the location in the lattice.
Due to the presence of $w(n)$ in (1.10), we call it
the general Jacobi system, whereas the Jacobi system [1,8,9,28,29]
corresponds to the case $w(n)\equiv 1.$

The system given in (1.10) is used as a model for various
physical systems. It describes the behavior of particles in
a one-dimensional lattice where each particle may experience
a local force as well as a force from the nearest neighbors.
By including the weight factors $w(n),$ we can use (1.10)
to describe wave propagation in a lattice where the
propagation speed may depend on the location.

The coefficients $a(n),$ $b(n),$ $w(n)$ appearing
in (1.10) are assumed to belong to
class $\Cal A$ specified below. The resulting restrictions
on the coefficients allow us to develop a mathematical scattering
theory for (1.10) and to assure the finiteness of the number of bound
states associated with (1.10).

\noindent {\bf Definition 1.1} {The coefficients $a(n),$ $b(n),$ $w(n)$
belong to class $\Cal A$ if they satisfy the following properties:}

\item{(a)} {\it They are real valued, and
$a(n)\ne 0$ and $w(n)>0$ for $n\in \bold Z.$}

\item{(b)} {\it They have the finite limits
$a_\infty,$ $b_\infty,$ $w_\infty,$ respectively, as $n\to \pm\infty,$
and $a_\infty\ne 0$ and $w_\infty>0.$}

\item{(c)} {\it They satisfy the restriction}
$$\ds\sum_{n=-\infty}^{+\infty} |n|\left(
\left| \ds\frac{a(n+1)}{\sqrt{w(n)\,w(n+1)}}-\ds\frac{a_\infty}{w_\infty}\right|
+\left| \ds\frac{b(n)}{w(n)}-\ds\frac{b_\infty}{w_\infty}\right|
\right)<+\infty.\tag 1.11$$

We remark that the restriction (1.11) is the analog
of (1.2). Such a restriction
can be obtained from the mathematical theory [14,15]
available for the case $w(n)\equiv 1.$

Our paper is organized as follows. In Section~2 we introduce
the alternate parameter $z$ related to the spectral parameter $\lambda$
via (2.1).
We introduce the Jost solutions from the left and from the right, respectively,
and also introduce
two other solutions related to the Jost solutions.
With the help of such solutions, the scattering
coefficients are obtained as functions of $z$ and their basic properties
are provided. In Section~3 we introduce the transition matrix
related to the scattering coefficients as in (3.1).
Next, we fragment the full-line lattice into $N$ pieces.
We then provide our fundamental result, namely,
the factorization formula (3.5) expressing the relationship
between the transition matrix for the full-line lattice and
the transition matrices for the fragments. The proof of the factorization
formula is first given when $N=2$ and then extended to an arbitrary
number of fragments using mathematical induction.

\vskip 10 pt
\noindent {\bf 2. THE GENERAL JACOBI SYSTEM AND SCATTERING COEFFICIENTS}
\vskip 3 pt

In this section, in preparation for the
derivation of the analog of the factorization formula (1.9),
we introduce the scattering coefficients for the general
Jacobi system (1.10)
and present their relevant properties that are needed later on.

Instead of using the spectral parameter
$\lambda$ in (1.10), we can equivalently use the parameter
$z,$ which is related to $\lambda$ as
$$\lambda=\ds\frac{a_\infty\,(z+z^{-1})+b_\infty}{w_\infty}.\tag 2.1$$
Let us define
$$\lambda_{\text{min}}:=\ds\frac{-2\,|a_\infty|+b_\infty}{w_\infty},
\quad
\lambda_{\text{max}}:=\ds\frac{2\,|a_\infty|+b_\infty}{w_\infty}.$$
When $a_\infty<0,$ the transformation $\lambda\mapsto z$ in (2.1)
maps the real $\lambda$-axis to the boundary of the upper half
of the unit disc in such a way that
the interval
$\lambda\in(-\infty,\lambda_{\text{min}})$
is mapped to the real interval
$z\in (0,1),$ the
real interval $\lambda\in(\lambda_{\text{min}},\lambda_{\text{max}})$
is mapped to $z=e^{i\theta}$ with
$\theta\in(0,\pi),$
the real interval $\lambda\in(\lambda_{\text{max}},+\infty)$
is mapped to the real interval $z\in(-1,0),$
while $\lambda=\lambda_{\text{min}}$ is mapped to $z=1$ and
$\lambda=\lambda_{\text{max}}$ is mapped to $z=-1.$
On the other hand, when $a_\infty>0,$ the transformation $\lambda\mapsto z$
maps the real $\lambda$-axis to the boundary of the lower half
of the unit disc in such a way that
the interval
$\lambda\in(-\infty,\lambda_{\text{min}})$
is mapped to the real interval
$z\in (-1,0),$ the
real interval $\lambda\in(\lambda_{\text{min}},\lambda_{\text{max}})$
is mapped to $z=e^{i\theta}$ with
$\theta\in(-\pi,0),$
the real interval $\lambda\in(\lambda_{\text{max}},+\infty)$
is mapped to the real interval $z\in(0,1),$
while $\lambda=\lambda_{\text{min}}$ is mapped to $z=-1$ and
$\lambda=\lambda_{\text{max}}$ is mapped to $z=1.$

Using (2.1) in (1.10), we can write (1.10) in terms of the parameter $z$ as
$$a(n+1)\,\phi(z,n+1)+b(n)\,\phi(z,n)+
a(n)\,\phi(z,n-1)=
\ds\frac{w(n)}{w_\infty}\,\left[a_\infty(z+z^{-1})+b_\infty\right]\phi(z,n),
\tag 2.2$$
where $n\in\bold Z$ and the $z$-values now
occur on the unit circle $\bold T$ in the complex $z$-plane
given by $|z|=1.$
 The unperturbed equation corresponding to
 (2.2) is obtained by replacing
 $a(n),$ $b(n),$ $w(n)$ with their limiting values
 $a_\infty,$ $b_\infty,$ $w_\infty,$ respectively, and we have
$$\overset{\circ}\to\phi(z,n+1)+
\overset{\circ}\to\phi(z,n-1)=
(z+z^{-1})\,\overset{\circ}\to\phi(z,n),
\qquad n\in\bold Z.
\tag 2.3$$

The difference equation (2.2) with the coefficients
belonging to class $\Cal A$
specified in Definition~1.1 has [9] two linearly independent solutions, namely
the Jost solution from the left $f_{\text l}(z,n)$ satisfying
$$f_{\text l}(z,n)=z^n \left[1+o(1)\right],\qquad
n\to +\infty,\tag 2.4$$
and the Jost solution from the right $f_{\text r}(z,n)$ satisfying
$$f_{\text r}(z,n)=z^{-n} \left[1+o(1)\right],\qquad
n\to -\infty.\tag 2.5$$
The scattering coefficients $T,$ $R,$ $L$ are now functions of
the variable $z,$ and they are obtained
as in (1.5) and (1.6) from the spatial asymptotics of the Jost
solutions as
$$f_{\text l}(z,n)=\ds\frac{1}{T(z)}\,z^n +\ds\frac{L(z)}{T(z)}\,z^{-n}
+o(1),\qquad
n\to -\infty,\tag 2.6$$
$$f_{\text r}(z,n)=\ds\frac{1}{T(z)}\,z^{-n} +\ds\frac{R(z)}{T(z)}\,z^n
+o(1),\qquad
n\to +\infty.\tag 2.7$$

There are two other solutions to (2.2) related to the
Jost solutions. Let us use $g_{\text l}(z,n)$ and $g_{\text r}(z,n)$ to denote them
and introduce them as
$$g_{\text l}(z,n):=f_{\text l}(z^{-1},n),
\quad g_{\text r}(z,n):=f_{\text r}(z^{-1},n).\tag 2.8$$
 Since $z$ and $z^{-1}$ appear symmetrically in (2.2), it follows that
$g_{\text l}(z,n)$ and $g_{\text r}(z,n)$ satisfy (2.2). Using
(2.8) in (2.4) and (2.5),
it follows that $g_{\text l}(z,n)$ and $g_{\text r}(z,n)$
satisfy the respective asymptotics
$$g_{\text l}(z,n)=z^{-n} \left[1+o(1)\right],\qquad
n\to +\infty,\tag 2.9$$
$$g_{\text r}(z,n)=z^n \left[1+o(1)\right],\qquad
n\to -\infty.\tag 2.10$$

The Wronskian $[\phi(z,n);\zeta(z,n)]$ of any two solutions $\phi(z,n)$ and
$\zeta(z,n)$ to (2.2) is defined [9] as
$$[\phi(z,n);\zeta(z,n)]:=a(n+1)\big( \phi(z,n)\, \zeta(z,n+1)-
\phi(z,n+1)\, \zeta(z,n)\big),\tag 2.11$$
and it is known [9] that the value of the Wronskian is independent of
$n$ and hence can be evaluated at any $n$-value or as $n\to\pm\infty.$
We have introduced the scattering coefficients through the spatial asymptotics
of the Jost solutions via (2.6) and (2.7). Alternatively,
the scattering coefficients can be obtained by using some Wronskians
involving
$f_{\text l}(z,n),$ $f_{\text r}(z,n),$ $g_{\text l}(z,n),$ and $g_{\text r}(z,n).$
It is possible to determine the basic properties of the scattering
coefficients by evaluating such Wronskians as $n\to+\infty$
and also as $n\to-\infty$ and by equating the resulting expressions.

\noindent {\bf Theorem 2.1} {\it Assume that
the coefficients in (2.2) belong to class $\Cal A.$
Let $f_{\text l}(z,n)$ and $f_{\text r}(z,n)$
be the Jost solutions from the left and from the right, respectively,
to (2.2), and let $T(z),$ $R(z),$ $L(z)$ be the
 corresponding scattering
 coefficients appearing in (2.6) and (2.7). Then:}

\item{(a)} {\it The $z$-domain
of the Jost solutions $f_{\text l}(z,n)$ and $f_{\text r}(z,n)$
can be extended to $z\in\bold T$ with the help of}
$$f_{\text l}(z^{-1},n)=f_{\text l}(z^\ast,n)=f_{\text l}(z,n)^\ast,
\quad f_{\text r}(z^{-1},n)=f_{\text r}(z^\ast,n)=f_{\text r}(z,n)^\ast,
\qquad z\in\bold T,\tag 2.12$$
{\it where the asterisk denotes complex conjugation.}

\item{(b)} {\it The $z$-domain
of the scattering coefficients is
$z\in\bold T,$ and we have for $z\in\bold T$}
$$T(z^{-1})=T(z^\ast)=T(z)^\ast, \quad R(z^{-1})=R(z^\ast)=R(z)^\ast, \quad
L(z^{-1})=L(z^\ast)=L(z)^\ast.\tag 2.13$$
{\it Furthermore, for
$z\in\bold T$ the scattering coefficients satisfy}
$$\ds\frac{1}{T(z)\, T(z^{-1})}-\ds\frac{L(z)\,L(z^{-1})}{T(z)\,T(z^{-1})}=1,\tag 2.14$$
$$\ds\frac{1}{T(z)\, T(z^{-1})}-\ds\frac{R(z)\,R(z^{-1})}{T(z)\,T(z^{-1})}=1,\tag 2.15$$
$$\ds\frac{R(z^{-1})}{T(z^{-1})}=-\ds\frac{L(z)}{T(z)},\quad
\ds\frac{L(z^{-1})}{T(z^{-1})}=-\ds\frac{R(z)}{T(z)},\tag 2.16$$
$$T(z)^2-R(z)\,L(z)=\ds\frac{T(z)}{T(z^{-1})}.\tag 2.17$$

\noindent PROOF: As stated in Definition~1.1, the coefficients
$a(n),$ $b(n),$ $w(n)$ and their limiting values
$a_\infty,$ $b_\infty,$ $w_\infty$ are all real valued.
For $z\in\bold T,$ we have $z^{-1}=z^\ast.$ When $z\in\bold T,$
replacing $z$ by $z^\ast$ in (2.2) and then taking the complex conjugate of
both sides of the resulting equation, we see that
$f_{\text l}(z^\ast,n)^\ast$ remains a solution to (2.2).
Furthermore, $f_{\text l}(z^\ast,n)^\ast$ satisfies the asymptotics
given in (2.4), and hence $f_{\text l}(z^\ast,n)^\ast=
f_{\text l}(z,n)$ when $z\in\bold T.$
By a similar argument we get
$f_{\text r}(z^\ast,n)^\ast=
f_{\text r}(z,n)$ when $z\in\bold T.$
Hence, (2.12) is proved.
Using (2.12) in (2.6) and (2.7) and the fact that
$z^{-1}=z^\ast$ for $z\in\bold T,$ we obtain
(2.13).
We get (2.14) by evaluating the Wronskian
$[f_{\text l}(z,n);g_{\text l}(z,n)]$ as $n\to +\infty$ and also
as $n\to-\infty$ and by equating the resulting expressions, 
where $g_{\text l}(z,n)$ is the quantity appearing in (2.8)
and (2.9). For this purpose,
using (2.4) and (2.9) in (2.11), we get the value of $[f_{\text l}(z,n);g_{\text l}(z,n)]$
as $n\to+\infty$ as
$$[f_{\text l}(z,n);g_{\text l}(z,n)]=a_\infty\,\left(z^{-1}-z\right).\tag 2.18$$
The same Wronskian, as $n\to-\infty,$ is evaluated with the help
of (2.6) and the first equality in (2.8) as
$$[f_{\text l}(z,n);g_{\text l}(z,n)]=a_\infty\,\left(z^{-1}-z\right)
\left(\ds\frac{1}{T(z)\, T(z^{-1})}-\ds\frac{L(z)\,L(z^{-1})}{T(z)\,T(z^{-1})}\right).\tag 2.19$$
Comparing (2.18) and (2.19) we establish (2.14).
Similarly, we obtain (2.15) by evaluating the Wronskian
$[f_{\text r}(z,n);g_{\text r}(z,n)]$ as $n\to\pm\infty$
and equating the resulting expressions.
We get the first equality in (2.16)
by evaluating the Wronskian
$[f_{\text l}(z,n);g_{\text r}(z,n)]$ as $n\to\pm\infty$
and equating the resulting expressions,
where $g_{\text r}(z,n)$ is the quantity appearing in (2.8)
and (2.10).
The second equality in (2.16) is obtained by evaluating the Wronskian
$[f_{\text r}(z,n);g_{\text l}(z,n)]$ as $n\to\pm\infty$
and equating the resulting expressions.
Finally, we get (2.17) by using the second equality of (2.16) in (2.14). \qed

\vskip 10 pt
\noindent {\bf 3. THE FACTORIZATION FORMULA FOR
THE GENERAL JACOBI SYSTEM}
\vskip 3 pt

Our goal in this section is to establish the analog of (1.9)
for the general Jacobi system (2.2). This will be accomplished
by fragmenting the full-line lattice $\bold Z$ and relating
the transition matrix for the entire lattice to the
transition matrices associated with the fragments.
The resulting factorization formula with two fragments is given
in (3.6), and the factorization formula for an arbitrary number of
fragments is given in (3.5).

For (2.2) with
coefficients $\{a(n),b(n),w(n)\}_{n\in\bold Z},$ we define
the transition matrix $\Lambda(z)$ for $z\in\bold T$
as
$$\Lambda(z):=\bm \ds\frac{1}{T(z)}& -\ds\frac{R(z)}{T(z)}\\
\stretch
\ds\frac{L(z)}{T(z)}& \ds\frac{1}{T(z^{-1})}\endbm,\tag 3.1$$
which resembles the transition matrix given in (1.7)
for the Schr\"odinger equation.

Let us partition the set of integers $\bold Z$ into $N$
ordered subsets as
$$\bold Z=\{n_0,\dots,n_1\}\cup \{n_1+1,\dots,n_2\} \cup \cdots \cup \{n_{N-1}+1,\dots,n_N\},$$
where $N\ge 2$ and we have
$$n_0:=-\infty,\quad n_1<n_2<\cdots < n_{N-1},\quad n_N:=+\infty.$$
We then obtain the fragmentation of the coefficients in
(1.10) into $N$ groups as
$$\left(a(n),b(n),w(n)\right)=\cases
\left(a_1(n),b_1(n),w_1(n)\right),\qquad & n_0<n\le n_1,\\
\stretch
\left(a_2(n),b_2(n),w_2(n)\right),\qquad & n_1<n\le n_2,\\
\quad \vdots &\quad \vdots  \\
\left(a_{N-1}(n),b_{N-1}(n),w_{N-1}(n)\right),\qquad & n_{N-2}<n\le n_{N-1},\\
\stretch
\left(a_N(n),b_N(n),w_N(n)\right),\qquad & n_{N-1}<n\le n_N\endcases\tag 3.2$$
where for $j=1,2,\dots,N$ we have defined
$$\cases a_j(n):=a(n),\quad b_j(n):=b(n),\quad
w_j(n):=w(n), &\qquad n_{j-1}<n\le n_j,\\
\stretch
a_j(n):=a_\infty,\quad b_j(n):=b_\infty,\quad
w_j(n):=w_\infty, &\qquad n\le n_{j-1} \,\text{ or }\, n> n_j.\endcases\tag 3.3$$

For each fixed $j,$
let us use $f_{{\text l}j}(z,n)$ and $f_{{\text r}j}(z,n)$ to denote the
Jost solutions from the left and from the right, respectively,
corresponding to the coefficients $\{a_j(n),b_j(n),w_j(n)\}_{n\in\bold Z}.$ Let us also use
$T_j,$ $R_j,$ $L_j$ for the respective scattering coefficients corresponding
to $\{a_j(n),b_j(n),w_j(n)\}_{n\in\bold Z}.$
Let $\Lambda_j(z)$ be the corresponding transition matrix
for $\{a_j(n),b_j(n),w_j(n)\}_{n\in\bold Z}$ defined in a similar way as in (3.1) as
$$\Lambda_j(z):=\bm \ds\frac{1}{T_j(z)}& -\ds\frac{R_j(z)}{T_j(z)}\\
\stretch
\ds\frac{L_j(z)}{T_j(z)}& \ds\frac{1}{T_j(z^{-1})}\endbm.\tag 3.4$$

Corresponding to the fragmentation (3.2), we are interested in
proving the factorization formula
$$\Lambda(z)=\Lambda_1(z)\, \Lambda_2(z)\, \cdots \Lambda_N(z),
\qquad z\in\bold T,\tag 3.5$$
where the right-hand side consists of the product of the $2\times 2$
transition matrices
in the ordered indicated.
It is enough to prove (3.5) when
the number of fragments is 2, i.e. $N=2$ in (3.5),
because the case $N>2$
can be obtained via mathematical induction. Thus, we would like to prove that
$$\bm \ds\frac{1}{T(z)}& -\ds\frac{R(z)}{T(z)}\\
\stretch
\ds\frac{L(z)}{T(z)}& \ds\frac{1}{T(z^{-1})}\endbm=
\bm\ds\frac{1}{T_1(z)}& -\ds\frac{R_1(z)}{T_1(z)}\\
\stretch
\ds\frac{L_1(z)}{T_1(z)}& \ds\frac{1}{T_1(z^{-1})}\endbm
\bm
\ds\frac{1}{T_2(z)}& -\ds\frac{R_2(z)}{T_2(z)}\\
\stretch
\ds\frac{L_2(z)}{T_2(z)}& \ds\frac{1}{T_2(z^{-1})}\endbm,
\qquad z\in\bold T.\tag 3.6$$

In order to prove the factorization formula with two fragments, i.e. to prove
the formula given in (3.6), we need a series
of auxiliary results presented in the next several propositions.

In our first proposition,
when $N=2$ in the partitioning in (3.2),
we express the Jost solutions $f_{\text l}(z,n)$
and $f_{\text r}(z,n)$ in terms of the Jost solution
$f_{{\text l}2}(z,n)$ and its relative
$g_{{\text l}2}(z,n)$ for $n\ge n_1.$
The result is needed later on in the proof of the factorization formula
(3.6).

\noindent {\bf Proposition 3.1} {\it Assume that
the coefficients in (2.2) belong to class $\Cal A.$
Further assume that $N=2$ in the partitioning given in (3.2).
Let $f_{\text l}(z,n)$ and $f_{\text r}(z,n)$
be the Jost solutions from the left and from the right, respectively,
to (2.2), and let $f_{{\text l}2}$ be the Jost solution
 from the left for (2.2) with the coefficients
 $\{a_2(n),b_2(n),w_2(n)\}_{n\in\bold Z}$ specified in (3.3). Let
$g_{{\text l}2}(z,n)$ be the quantity related to $f_{{\text l}2}(z,n)$
as in the first equality in (2.8), namely
$$g_{{\text l}2}(z,n)\equiv f_{{\text l}2}(z^{ -1},n).\tag 3.7$$
Then, for $n\ge n_1$ we have}
$$\bm f_{\text l}(z,n)& f_{\text r}(z,n)\\
\stretch
f_{\text l}(z,n+1)& f_{\text r}(z,n+1)\endbm=
\bm f_{{\text l}2}(z,n)& \alpha(z)\,f_{{\text l}2}(z,n)
+\beta(z)\,g_{{\text l}2}(z,n)\\
\stretch
f_{{\text l}2}(z,n+1)& \alpha(z)\,
f_{{\text l}2}(z,n+1)+\beta(z)\,g_{{\text l}2}(z,n+1)\endbm,
\tag 3.8$$
{\it where}
$$\alpha(z)=\ds\frac{R(z)}{T(z)},\quad \beta(z)=\ds\frac{1}{T(z)}.\tag 3.9$$

\noindent PROOF: Because of the second line of (3.2),
it follows that, for $n\ge n_1+1,$ both $f_{\text l}(z,n)$
and $f_{{\text l}2}(z,n)$ satisfy the same equation, namely (2.2), and the
same asymptotic condition, namely (2.4). Thus, we obtain
$$f_{\text l}(z,n)=f_{{\text l}2}(z,n),\qquad n\ge n_1+1.$$
Because $z$ and $z^{-1}$ appear symmetrically in (2.2),
it follows from (3.7) that
 $f_{{\text l}2}(z,n)$ and $g_{{\text l}2}(z,n)$ both satisfy
 (2.2) with $n\ge n_1+1,$ and hence $f_{\text r}(z,n)$ can be expressed
 as a linear combination of $f_{{\text l}2}(z,n)$ and $g_{{\text l}2}(z,n)$
 for $n\ge n_1+1.$ Thus, (3.8) holds for $n\ge n_1$ provided we can prove that
$$f_{\text l}(z,n_1)=f_{{\text l}2}(z,n_1),
\quad  f_{\text r}(z,n_1)=\alpha(z)\,f_{{\text l}2}(z,n_1)
+\beta(z)\,g_{{\text l}2}(z,n_1).\tag 3.10$$
Using (2.2) at $n=n_1+1$ with $f_{\text l}(z,n)$
and $\{a(n),b(n),w(n)\}_{n\in\bold Z}$ and also
using (2.2) at $n=n_1+1$ with $f_{{\text l}2}(z,n)$ and
$\{a_2(n),b_2(n),w_2(n)\}_{n\in\bold Z},$
we obtain
$$a(n_1+1)\,f_{\text l}(z,n_1)=a_2(n_1+1)\,f_{{\text l}2}(z,n_1),$$
and hence, from the facts that $a(n_1+1)=a_2(n_1+1)$ and
$a(n_1+1)>0,$ we get the first equality in (3.10).
Proceeding similarly, we get the second equality in (3.10).
By letting $n\to+\infty$ on both sides of
(3.8) and using (2.4), (2.7), and (2.9) we get (3.9). \qed

In the next proposition,
when $N=2$ in the partitioning in (3.2),
we express the Jost solutions $f_{\text l}(z,n)$
and $f_{\text r}(z,n)$ in terms of the Jost solution
$f_{{\text r}1}(z,n)$ and its relative
$g_{{\text r}1}(z,n)$ for $n\le n_1.$
The result is needed to prove the factorization formula
given in (3.6).

\noindent {\bf Proposition 3.2} {\it Assume that
the coefficients in (2.2) belong to class $\Cal A.$
Further assume that $N=2$ in the partitioning given in (3.2).
Let $f_{\text l}(z,n)$ and $f_{\text r}(z,n)$
be the Jost solutions from the left and from the right, respectively,
to (2.2), and let $f_{{\text r}1}$ be the Jost solution
 from the right for (2.2) with the coefficients
 $\{a_1(n),b_1(n),w_1(n)\}_{n\in\bold Z}$ given in (3.3). Let
$g_{{\text r}1}(z,n)$ be the quantity related to $f_{{\text r}1}(z,n)$
as in the second equality in (2.8), namely}
$$g_{{\text r}1}(z,n)\equiv f_{{\text r}1}(z^{-1},n).\tag 3.11$$
{\it Then, for $n\le n_1-1,$ we have}
$$\bm f_{\text l}(z,n)& f_{\text r}(z,n)\\
\stretch
f_{\text l}(z,n+1)& f_{\text r}(z,n+1)\endbm=
\bm \gamma(z)\,g_{{\text r}1}(z,n)+\epsilon(z)\,f_{{\text r}1}(z,n)&
f_{{\text r}1}(z,n) \\
\stretch
\gamma(z)\,g_{{\text r}1}(z,n+1)+\epsilon(z)\,f_{{\text r}1}(z,n+1)&
f_{{\text r}1}(z,n+1)\endbm,\tag 3.12$$
{\it and we also have}
$$f_{\text r}(z,n_1+1)=\ds\frac{a_\infty}{a(n_1+1)}\,
f_{{\text r}1}(z,n_1+1),\tag 3.13$$
$$f_{\text l}(z,n_1+1)=\ds\frac{a_\infty}{a(n_1+1)}\left[
\gamma(z)\,g_{{\text r}1}(z,n_1+1)+\epsilon(z)\,f_{{\text r}1}(z,n_1+1)\right],
\tag 3.14$$
{\it where}
$$\gamma(z)=\ds\frac{L(z)}{T(z)},\quad \epsilon(z)=\ds\frac{1}{T(z)}.\tag 3.15$$

\noindent PROOF: Because of the first line of (3.2),
it follows that, for $n\le n_1-1,$ both $f_{\text r}(z,n)$
and $f_{{\text r}1}(z,n)$ satisfy the same equation, namely (2.2), and the
same asymptotic condition, namely (2.5). Thus, we obtain
$$f_{\text r}(z,n)=f_{{\text r}1}(z,n),\qquad n\le n_1-1.\tag 3.16$$
 From (2.2) with $n=n_1-1$ it then follows that (3.16) actually holds for $n=n_1$
 as well and we obtain
$$f_{\text r}(z,n)=f_{{\text r}1}(z,n),\qquad n\le n_1.$$
 From the second equation in (2.8), it follows that
 $f_{{\text r}1}(z,n)$ and $g_{{\text r}1}(z,n)$ both satisfy
 (2.2) with $n\le n_1-1,$ and hence $f_{\text l}(z,n)$ can be expressed
 as a linear combination of $f_{{\text r}1}(z,n)$ and $g_{{\text r}1}(z,n)$
 for $n\le n_1,$ which is proved by proceeding in a similar way as in the proof of
 Proposition~3.1.
In a similar way,
using (2.2) with $n=n_1$ and the facts that we have (3.16), $a(n)>0,$ and $a_1(n_1+1)=a_\infty,$
 we obtain (3.13). Similarly, using (2.2) with
 $n=n_1$ and making use of the equality of the $(1,1)$-entries in (3.12) for
 $n=n_1,$ we obtain (3.14).
 Finally, letting $n\to-\infty$ in (3.12) and using (2.5),
 (2.6), and (2.10) we obtain
 (3.15). \qed

The result in the next proposition is needed in the proof of Theorem~3.6.

\noindent {\bf Proposition 3.3} {\it Assume that
the coefficients in (2.2) belong to class $\Cal A.$
Further assume that $N=2$ in the partitioning given in (3.2).
Let $f_{{\text l}2}(z,n)$ be the Jost solution
 from the left for (2.2) with the coefficients
 $\{a_2(n),b_2(n),w_2(n)\}_{n\in\bold Z}$ given in (3.3).
 Let $T_2(z),$ $R_2(z),$ $L_2(z)$ be the scattering
 coefficients associated with the coefficients
 $\{a_2(n),b_2(n),w_2(n)\}_{n\in\bold Z}.$ Then, we have}
$$f_{{\text l}2}(z,n)=\ds\frac{1}{T_2(z)}\,z^n+\ds\frac{L_2(z)}{T_2(z)}\,z^{-n},
\qquad n\le n_1.\tag 3.17$$
$$f_{{\text l}2}(z,n_1+1)=\ds\frac{a_\infty}
{a(n_1+1)}\left[\ds\frac{1}{T_2(z)}\,z^{n_1+1}
+\ds\frac{L_2(z)}{T_2(z)}\,z^{-n_1-1}\right].\tag 3.18$$

\noindent PROOF: When $n\le n_1-1$ in (2.2), the Jost solution
$f_{{\text l}2}(z,n)$ satisfies (2.3) with $n\le n_1-1.$ Furthermore,
from the analog of (2.6) for $f_{{\text l}2}(z,n)$ we already have
(3.17) for $n\le n_1-1.$ Then, (3.2) with $n=n_1-1$
implies that (3.17) actually holds for $n=n_1$ as well. Thus, (3.17)
is proved. Then, using (2.2) with $n=n_1$ and utilizing
(3.17) for $n=n_1$ and $n=n_1-1,$
after some simplifications, we obtain (3.18). \qed

We need the following analog of Proposition~3.3, which is needed in the
proof of the factorization formula given in (3.6).

\noindent {\bf Proposition 3.4} {\it Assume that
the coefficients in (2.2) belong to class $\Cal A.$
Further assume that $N=2$ in the partitioning given in (3.2).
Let $f_{{\text r}1}$ be the Jost solution
 from the right for (2.2) with the coefficients
 $\{a_1(n),b_1(n),w_1(n)\}_{n\in\bold Z}$ appearing in (3.3).
 Let $T_1(z),$ $R_1(z),$ $L_1(z)$ be the scattering
 coefficients associated with the coefficients
 $\{a_1(n),b_1(n),w_1(n)\}_{n\in\bold Z}.$ Then, we have}
$$f_{{\text r}1}(z,n)=\ds\frac{1}{T_1(z)}\,z^{-n}+\ds\frac{R_1(z)}{T_1(z)}\,z^n,
\qquad n\ge n_1.\tag 3.19$$

\noindent PROOF:
When $n\ge n_1+1$ in (2.2), the Jost solution
$f_{{\text r}1}(z,n)$ satisfies (2.3) with $n\ge n_1+1.$ Furthermore,
from the analog of (2.7) for $f_{{\text r}1}(z,n)$ we already have
(3.19) for $n\ge n_1+1.$ Then, (2.3) with $n=n_1+1$
implies that (3.19) actually holds for $n=n_1$ as well. \qed

The next proposition is needed in the proof of Theorem~3.6.

\noindent {\bf Proposition 3.5} {\it Assume that
the coefficients in (2.2) belong to class $\Cal A.$
Further assume that $N=2$ in the partitioning given in (3.2).
Let $f_{{\text r}1}$ be the Jost solution
 from the right for (2.2) with the coefficients
 $\{a_1(n),b_1(n),w_1(n)\}_{n\in\bold Z}$ appearing in (3.3).
 Let $T_1(z),$ $R_1(z),$ $L_1(z)$ be the scattering
 coefficients associated with the coefficients
 $\{a_1(n),b_1(n),w_1(n)\}_{n\in\bold Z}.$
Similarly, let $f_{{\text l}2}$ be the Jost solution
 from the left for (2.2) with the coefficients
 $\{a_2(n),b_2(n),w_2(n)\}_{n\in\bold Z}$ appearing in (3.3).
 Let $T_2(z),$ $R_2(z),$ $L_2(z)$ be the scattering
 coefficients associated with the coefficients
 $\{a_2(n),b_2(n),w_2(n)\}_{n\in\bold Z}.$
 Furthermore, let $g_{{\text r}1}$ and
 $g_{{\text l}2}$ be the quantities appearing in (3.11) and
 (3.7), respectively.
 Then, we have}
$$\bm f_{{\text l}2}(z,n_1)& g_{{\text l}2}(z,n_1)\\
\stretch
f_{{\text l}2}(z,n_1+1)& g_{{\text l}2}(z,n_1+1)\endbm=
\bm 1& 0\\
\stretch
0& \ds\frac{a_\infty}{a(n_1+1)}
\endbm
\bm \ds z^{n_1}& \ds  z^{-n_1}\\
\stretch
\ds  z^{n_1+1}& \ds  z^{-n_1-1}
\endbm
\bm \ds\frac{1}{T_2(z)}& \ds\frac{L_2(z^{-1})}{T_2(z^{-1})}\\
\stretch
\ds\frac{L_2(z)}{T_2(z)}& \ds\frac{1}{T_2(z^{-1})}
\endbm,\tag 3.20$$
$$\bm g_{{\text r}1}(z,n_1)& f_{{\text r}1}(z,n_1)\\
\stretch
g_{{\text r}1}(z,n_1+1)& f_{{\text r}1}(z,n_1+1)\endbm=
\bm \ds z^{n_1}& \ds  z^{-n_1}\\
\stretch
\ds  z^{n_1+1}& \ds  z^{-n_1-1}
\endbm
\bm \ds\frac{1}{T_1(z^{-1})}& \ds\frac{R_1(z)}{T_1(z)}\\
\stretch
\ds\frac{R_1(z^{-1})}{T_1(z^{-1})}& \ds\frac{1}{T_1(z)}
\endbm.\tag 3.21$$

\noindent PROOF: We obtain (3.20) by using (3.7), (3.17), and (3.18).
Similarly, we get (3.21) by using (3.11) and (3.19). \qed

In the next theorem, we prove the factorization formula when
there are two fragments. The factorization formula (3.5) with $N$ fragments
can then be proved via mathematical induction.

\noindent {\bf Theorem 3.6} {\it Assume that
the coefficients in (2.2) belong to class $\Cal A.$
Further assume that $N=2$ in the partitioning given in (3.2).
Let $f_{{\text r}1}$ be the Jost solution
 from the right for (2.2) with the coefficients
 $\{a_1(n),b_1(n),w_1(n)\}_{n\in\bold Z}$ appearing in (3.3).
 Let $T_1(z),$ $R_1(z),$ $L_1(z)$ be the scattering
 coefficients associated with the coefficients
 $\{a_1(n),b_1(n),w_1(n)\}_{n\in\bold Z}.$
Similarly, let $f_{{\text l}2}$ be the Jost solution
 from the left for (2.2) with the coefficients
 $\{a_2(n),b_2(n),w_2(n)\}_{n\in\bold Z}$ appearing in (3.3).
 Let $T_2(z),$ $R_2(z),$ $L_2(z)$ be the scattering
 coefficients associated with the coefficients
 $\{a_2(n),b_2(n),w_2(n)\}_{n\in\bold Z}.$
 Furthermore, let $g_{{\text r}1}$ and
 $g_{{\text l}2}$ be the quantities appearing in (3.11) and
 (3.7), respectively.
 Then, the factorization formula given in (3.6) holds.}

 \noindent PROOF: We will derive (3.6) by evaluating
 the left-hand side of (3.8) when $n=n_1$ in two
 different ways and by equating the resulting expressions.
 The first expression will be obtained by using (3.8) and
 the second will be obtained by using (3.12). From (3.8) and (3.9)
 we get
$$\bm f_{\text l}(z,n_1)& f_{\text r}(z,n_1)\\
\stretch
f_{\text l}(z,n_1+1)& f_{\text r}(z,n_1+1)\endbm=
\bm f_{{\text l}2}(z,n_1)& g_{{\text l}2}(z,n_1)\\
\stretch
f_{{\text l}2}(z,n_1+1)& g_{{\text l}2}(z,n_1+1)\endbm
\bm 1& \ds\frac{R(z)}{T(z)}\\
\stretch
0& \ds\frac{1}{T(z)}\endbm.\tag 3.22$$
On the other hand, from (3.12)-(3.15) we get
$$\bm f_{\text l}(z,n_1)& f_{\text r}(z,n_1)\\
\stretch
f_{\text l}(z,n_1+1)& f_{\text r}(z,n_1+1)\endbm=\bm 1& 0\\
\stretch
0& \ds\frac{a_\infty}{a(n_1+1)}
\endbm
\bm g_{{\text r}1}(z,n_1)& f_{{\text r}1}(z,n_1)\\
\stretch
g_{{\text r}1}(z,n_1+1)& f_{{\text r}1}(z,n_1+1)\endbm
\bm \ds\frac{1}{T(z)}& 0\\
\stretch
 \ds\frac{L(z)}{T(z)}& 1\endbm.\tag 3.23$$
 Thus, the right-hand side of (3.22) must be equal to
 the right-hand side of (3.23).
Using (3.20) on the right-hand side of (3.22) and
using (3.21) on the right-hand side of (3.23)
and equating the resulting expressions, after some simplification, we obtain
$$\bm \ds\frac{1}{T_2(z)}& \ds\frac{L_2(z^{-1})}{T_2(z^{-1})}\\
\stretch
\ds\frac{L_2(z)}{T_2(z)}& \ds\frac{1}{T_2(z^{-1})}
\endbm
\bm 1& \ds\frac{R(z)}{T(z)}\\
\stretch
0& \ds\frac{1}{T(z)}\endbm=
\bm \ds\frac{1}{T_1(z^{-1})}& \ds\frac{R_1(z)}{T_1(z)}\\
\stretch
\ds\frac{R_1(z^{-1})}{T_1(z^{-1})}& \ds\frac{1}{T_1(z)}
\endbm
\bm \ds\frac{1}{T(z)}& 0\\
\stretch
 \ds\frac{L(z)}{T(z)}& 1\endbm,$$
 or equivalently we have
$$
\bm \ds\frac{1}{T_1(z^{-1})}& \ds\frac{R_1(z)}{T_1(z)}\\
\stretch
\ds\frac{R_1(z^{-1})}{T_1(z^{-1})}& \ds\frac{1}{T_1(z)}
\endbm^{-1}
\bm \ds\frac{1}{T_2(z)}& \ds\frac{L_2(z^{-1})}{T_2(z^{-1})}\\
\stretch
\ds\frac{L_2(z)}{T_2(z)}& \ds\frac{1}{T_2(z^{-1})}
\endbm
=
\bm \ds\frac{1}{T(z)}& 0\\
\stretch
 \ds\frac{L(z)}{T(z)}& 1\endbm
 \bm 1& \ds\frac{R(z)}{T(z)}\\
\stretch
0& \ds\frac{1}{T(z)}\endbm^{-1}.\tag 3.24$$
In a straightforward way we evaluate the inverse matrices appearing in (3.24) as
$$\bm 1& \ds\frac{R(z)}{T(z)}\\
\stretch
0& \ds\frac{1}{T(z)}\endbm^{-1}=
\bm 1& -R(z)\\
\stretch
0& \ds\frac{1}{T(z)}\endbm,\tag 3.25$$
$$\bm \ds\frac{1}{T_1(z^{-1})}& \ds\frac{R_1(z)}{T_1(z)}\\
\stretch
\ds\frac{R_1(z^{-1})}{T_1(z^{-1})}& \ds\frac{1}{T_1(z)}
\endbm^{-1}=\ds\frac{1}
{\ds\frac{1}{T_1(z)}\,\ds\frac{1}{T_1(z^{-1})}-
\ds\frac{R_1(z)}{T_1(z)}\ds\frac{R_1(z^{-1})}{T_1(z^{-1})}
}
\bm \ds\frac{1}{T_1(z)} & -\ds\frac{R_1(z)}{T_1(z)}\\
\stretch
-\ds\frac{R_1(z^{-1})}{T_1(z^{-1})}& \ds\frac{1}{T_1(z^{-1})}
\endbm.\tag 3.26$$
 From (2.15) we see that the determinantal quantity appearing
 as a coefficient on the right-hand side
 of (3.26) is equal to $1.$ Thus, (3.26) simplifies to
$$\bm \ds\frac{1}{T_1(z^{-1})}& \ds\frac{R_1(z)}{T_1(z)}\\
\stretch
\ds\frac{R_1(z^{-1})}{T_1(z^{-1})}& \ds\frac{1}{T_1(z)}
\endbm^{-1}=
\bm \ds\frac{1}{T_1(z)} & -\ds\frac{R_1(z)}{T_1(z)}\\
\stretch
-\ds\frac{R_1(z^{-1})}{T_1(z^{-1})}& \ds\frac{1}{T_1(z^{-1})}
\endbm.\tag 3.27$$
Using (3.25) and (3.27) in (3.24) we obtain
$$\bm \ds\frac{1}{T_1(z)} & -\ds\frac{R_1(z)}{T_1(z)}\\
\stretch
-\ds\frac{R_1(z^{-1})}{T_1(z^{-1})}& \ds\frac{1}{T_1(z^{-1})}
\endbm
\bm \ds\frac{1}{T_2(z)}& \ds\frac{L_2(z^{-1})}{T_2(z^{-1})}\\
\stretch
\ds\frac{L_2(z)}{T_2(z)}& \ds\frac{1}{T_2(z^{-1})}
\endbm
=
\bm \ds\frac{1}{T(z)}& 0\\
\stretch
 \ds\frac{L(z)}{T(z)}& 1\endbm
 \bm 1& -R(z)\\
\stretch
0& \ds\frac{1}{T(z)}\endbm
 .\tag 3.28$$
Using (2.16) on the left-hand side of (3.28) we see that the left-hand side
of (3.28) is equal to the matrix product
$\Lambda_1(z)\, \Lambda_2(z),$ where
$\Lambda_1(z)$ and $\Lambda_2(z)$ are the transition matrices defined in (3.4).
On the other hand, the right-hand side in (3.28) is equal to
the transition matrix $\Lambda(z)$ defined in (3.1). Thus,
(3.6) is established. \qed

Via mathematical induction,
the factorization formula (3.5) with $N$ fragments
holds, as stated in the next corollary.

\noindent {\bf Corollary 3.7} {\it Assume that
the coefficients in (2.2) belong to class $\Cal A.$
Further assume that we have the partitioning specified in (3.2)
 with the coefficients
 $\{a_j(n),b_j(n),w_j(n)\}_{n\in\bold Z}$ appearing in (3.3)
 for $1\le j\le N$ for any positive integer $N\ge 2.$
 Let $T_j(z),$ $R_j(z),$ $L_j(z)$ be the scattering
 coefficients associated with the coefficients
 $\{a_j(n),b_j(n),w_j(n)\}_{n\in\bold Z}.$
 Then, the factorization formula given in (3.5) holds,
 where $\Lambda(z)$ and $\Lambda_j(z)$ are
 the transition matrices defined in (3.1) and (3.4), respectively.}

\vskip 5 pt
\noindent {\bf Acknowledgments.} The first author expresses his gratitude to the Institute of Physics and Mathematics of the Universidad Michoacana de San Nicol\'as de
Hidalgo, M\'exico for its hospitality.
This work was completed with the support of the  Visiting
Distinguished Professor 2014 grant of
the Mexican Academy of Sciences.
The second author was partially supported by CIC-UMNSH and SNI-Conacyt.

\vskip 5 pt

\noindent {\bf{References}}

\item{[1]} N. Akhiezer,
{\it The classical moment problem,} Oliver and Boyd, London, 1965.

\item{[2]}
T. Aktosun, {\it A factorization of the scattering matrix for the
Schr\"odinger
equation and for the wave equation in one dimension,}
J. Math. Phys. {\bf 33}, 3865--3869 (1992).

\item{[3]}
T. Aktosun, {\it Factorization and small-energy asymptotics for the radial
Schr\"odinger
equation,}
J. Math. Phys. {\bf 41}, 4262--4270 (2000).

\item{[4]}
T. Aktosun, M. Klaus, and C. van der Mee, {\it  Factorization of scattering
matrices due to partitioning of potentials in one-dimensional Schr\"odinger-type equations,}
J. Math. Phys. {\bf 37}, 5897--5915 (1996).

\item{[5]}
T. Aktosun and P. E. Sacks, {\it Phase recovery with nondecaying potentials,}
Inverse Problems {\bf 16}, 821--838 (2000).

\item{[6]}
T. Aktosun and P. E. Sacks, {\it  Inversion of reflectivity data for nondecaying potentials,}
SIAM J. Appl. Math. {\bf 60}, 1340--1356 (2000).

\item{[7]}
T. Aktosun and P. E. Sacks, {\it Potential splitting and numerical solution of the inverse scattering problem on the line,}
Math. Methods Appl. Sci. {\bf 25}, 347--355 (2002).

\item{[8]} F. Atkinson,
{\it Discrete and continuous boundary problems,} Academic Press, New York,
1964.

\item{[9]}
K. M. Case and M. Kac,
{\it A discrete version of the inverse scattering problem,}
J. Math. Phys. {\bf 14}, 594--603 (1973).

\item{[10]} K. Chadan and P. C. Sabatier, {\it Inverse problems
    in quantum scattering theory,} 2nd ed., Springer, New York,
    1989.

\item{[11]} P. Deift and E. Trubowitz, {\it Inverse scattering
    on the line,} Commun. Pure Appl. Math. {\bf 32}, 121--251
    (1979).

\item{[12]} L. D. Faddeev, {\it Properties of the $S$-matrix of
    the one-dimensional Schr\"odinger equation,} Amer. Math.
    Soc. Transl. {\bf 65} (ser. 2), 139--166 (1967).

\item{[13]} I. M. Gel'fand and B. M. Levitan, {\it On the
determination of a differential equation from its spectral
function,} Amer. Math. Soc. Transl. {\bf 1} (ser. 2),
253--304 (1955).

\item{[14]} G. S. Guseinov, {\it The determination of an infinite Jacobi
matrix from the
scattering data,} Soviet Math. Dokl. {\bf 17}, 596--600 (1976).

\item{[15]} G. S. Guseinov, {\it The inverse problem of scattering theory for a second-order
difference equation on the whole axis,} Soviet Math. Dokl. {\bf 17}, 1684--1688
(1976).

\item{[16]}  K. A. Kiers and W. van Dijk,
{\it Scattering in one dimension: The coupled Schr\"odinger equation, threshold behaviour and Levinson’s theorem,}
J. Math. Phys. {\bf 32}, 6033--6058 (1999).

\item{[17]} V. Kostrykin and R. Schrader, {\it Kirchhoff's rule for quantum wires,}
J. Phys. A {\bf 37}, 595--630 (1996).

\item{[18]} V. Kostrykin and R. Schrader, {\it
Kirchhoff's rule for quantum wires. II: The inverse problem with possible applications to quantum computers,}
Fortschr. Phys. {\bf 48}, 703--716 (2000).

\item{[19]}
V. Kostrykin and R. Schrader,
{\it The generalized star product and the factorization of scattering matrices on graphs,}
J. Math. Phys. {\bf 42}, 1563--1598 (2001).

\item{[20]} B. M. Levitan, {\it
Inverse Sturm Liouville Problems,} VNU Science Press, Utrecht, 1987.

\item{[21]}
C. F. Majkrzak and N. F. Berk, {\it Exact determination of the phase in neutron reflectometry,}
Phys. Rev. B {\bf 52}, 10827--10830 (1995).

\item{[22]}
C. F. Majkrzak and N. F. Berk, {\it Exact determination of the phase in neutron reflectometry
by variation of the
surrounding medium,}
Phys. Rev. B {\bf 58}, 15416--15418 (1998).

\item{[23]}
C. F. Majkrzak, N. F. Berk, J. Dura,
S. K. Satija, A. Karim, J. Pedulla, and R. D. Deslattes,
{\it Direct inversion of
specular reflectometry,}
Phys. B {\bf 241--243}, 1101--1103 (1998).

\item{[24]} V. A. Marchenko,
{\it Sturm-Liouville operators and applications,}
Birkh\"auser, Basel, 1986.

\item{[25]} R. Redheffer,
{\it Difference equations and functional equations in transmission line theory,} in:
E. F. Beckenbach (ed.) {\it Modern mathematics
for the engineer,} 2nd series, McGraw-Hill, New York, 1961,
Ch. 12.

\item{[26]}
R. Redheffer,
{\it
On the relation of transmission-line theory to scattering and transfer,}
J. Math. and Phys.
{\bf 41}, 1--41 (1962).

\item{[27]} M. Sassoli de Bianchi and M. Di Ventra, {\it
Differential equations and factorization property for the one-dimensional
Schr\"odinger equation with position-dependent mass,} Eur. J. Phys. {\bf 16}, 260–-265 (1995).

\item{[28]} B. Simon,
 {\it The classical moment problem as a self-adjoint
  finite difference operator,} Adv. Math. {\bf 137}, 82--103 (1998).

\item{[29]}
G. Teschl,
{\it Jacobi operators and completely integrable nonlinear lattices,}
Amer. Math. Soc.,  Providence, RI, 2000.

\end